\documentclass[11pt]{article}
\usepackage{latexsym}
\usepackage{fullpage}

\usepackage{amsfonts}
\usepackage{multirow}
\def\Real{\mathbb{R}}
\def\Complex{\mathbb{C}}

\def\01{\{0,1\}}
\newcommand{\ceil}[1]{\lceil{#1}\rceil}

\newcommand{\ket}[1]{|#1\rangle}
\newcommand{\bra}[1]{\langle#1|}

\newcommand{\ev}[2]{\langle #1#2 \rangle} 

\newcommand{\Tr}{\mbox{\rm Tr}}
\newcommand{\norm}[1]{\mbox{$\parallel{#1}\parallel$}}
\newcommand{\diag}{\mbox{\rm diag}}

\newtheorem{theorem}{Theorem}

\newtheorem{claim}{Claim}

\newenvironment{proof}
{\noindent {\bf Proof. }}
{{\hfill $\Box$}\\
 \smallskip}

\newenvironment{sdp}[2]{
\smallskip
\begin{center}
\begin{tabular}{ll}
#1 & #2\\
subject to 
}
{
\end{tabular}
\end{center}
\smallskip
}

\newcommand{\mA}{\mathcal{A}}
\newcommand{\mB}{\mathcal{B}}
\newcommand{\paper}{paper}
\newcommand{\bSpace}{$$}
\newcommand{\eSpace}{$$}

\begin{document}
\title{Tsirelson bounds for generalized Clauser-Horne-Shimony-Holt inequalities}
\author{Stephanie Wehner\\
CWI, Kruislaan 413, 1098 SJ Amsterdam, The Netherlands\\
wehner@cwi.nl} 
\maketitle

\begin{abstract}
Quantum theory imposes a strict limit on the strength of non-local correlations.
It only allows for a violation of the CHSH inequality up to the value $2\sqrt{2}$, 
known as Tsirelson's bound. In this \paper, we consider generalized CHSH inequalities
based on many measurement settings with two possible measurement outcomes each. 
We demonstrate how to prove Tsirelson bounds for any such generalized CHSH inequality 
using semidefinite programming.
As an example, we show that for any shared entangled state and 
observables $X_1,\ldots,X_n$ and $Y_1,\ldots,Y_n$ 
with eigenvalues $\pm 1$ we have
$
|\langle X_1Y_1 \rangle + \langle X_2Y_1 \rangle + \langle X_2Y_2\rangle + \langle X_3Y_2 \rangle + 
\ldots + \langle X_nY_n \rangle - \langle X_1Y_n \rangle| \leq 2 n\cos\left(\pi/(2n)\right).
$
It is well known that there exist observables such that equality can be achieved. 
However, we show that these are indeed optimal. 
Our approach can easily be generalized to other inequalities for such observables.
\end{abstract}

Non-local correlations arise as the result of measurements performed on a quantum system
shared between two spatially separated parties. Imagine two parties, Alice and Bob,
who are given access to a shared quantum state $\ket{\Psi}$, but cannot communicate. In the simplest case, 
each of them is able to perform one of two possible measurements. Every measurement has two possible
outcomes labeled $\pm 1$.
Alice and Bob now measure $\ket{\Psi}$ using an independently chosen measurement setting and record
their outcomes. In order to obtain an accurate estimate for the correlation between their measurement
settings and the measurement outcomes, they perform this experiment many times using an identically
prepared state $\ket{\Psi}$ in each round.
Both classical and quantum theories impose limits on the strength of such non-local correlations.
In particular, both do not violate the non-signaling condition of special relativity. That is,
the local choice of measurement setting does not allow Alice and Bob to transmit information.
Limits on the strength of correlations which are possible in the framework of any
classical theory, i.e.~a framework based on local hidden variables, are known as 
\emph{Bell inequalities}~\cite{bell:epr}. The best known Bell inequality is the
Clauser, Horne, Shimony and Holt (CHSH) inequality~\cite{chsh:nonlocal}
\bSpace
|\ev{X_1}{Y_1} + \ev{X_1}{Y_2} + \ev{X_2}{Y_1} - \ev{X_2}{Y_2}| \leq 2,
\eSpace
where $\{X_1,X_2\}$ and $\{Y_1,Y_2\}$ are the observables representing the 
measurement settings of Alice and Bob respectively. $\ev{X_i}{Y_j} = \bra{\Psi}X_i \otimes Y_j\ket{\Psi}$
denotes the mean value of $X_i$ and $Y_j$.
Quantum mechanics allows for a violation of the CHSH inequality, but curiously still limits
the strength of nonlocal correlations. Tsirelson's bound~\cite{tsirel:original} says that for quantum mechanics
\bSpace
|\ev{X_1}{Y_1} + \ev{X_1}{Y_2} + \ev{X_2}{Y_1} - \ev{X_2}{Y_2}| \leq 2\sqrt{2}.
\eSpace

Peres demonstrated how to derive Bell inequalities~\cite{peres:allBell} even for more than two 
settings. As Froissart and Tsirelson~\cite{tsirel:hadron} have shown, these inequalities correspond to 
the faces of a polytope. Computing the boundary of the space of correlations that can be attained 
using a classical theory therefore corresponds to determining the faces of this polytope. However, 
determining bounds on the correlations that \emph{quantum} theory allows remains an even more difficult 
problem~\cite{massar:tsirel}. All Tsirelson's bounds are known for 
CHSH-type inequalities (also known as correlation inequalities) with two measurement settings and two outcomes for both Alice 
and Bob~\cite{tsirel:hadron}. Filipp and Svozil~\cite{filipp:bell} have considered the case of
three measurement settings analytically and conducted numerical studies for a larger number of settings.
Finally, Buhrman and Massar have shown a bound
for a generalized CHSH inequality using
three measurement settings with three outcomes each~\cite{massar:tsirel}.

In this \paper, we investigate the case where Alice and Bob can choose from $n$
measurement settings with two outcomes each. We use a completely different approach
based on semidefinite programming in combination with Tsirelson's seminal 
results~\cite{tsirel:original,tsirel:separated,tsirel:hadron}. This method is similar
to methods used in computer science for the two-way partitioning problem~\cite{boyd:book} 
and the approximation algorithm for MAXCUT by Goemans and Williamson~\cite{goemans:maxcut}. 
Cleve et al.~\cite{cleve:nonlocal}
have also remarked that Tsirelson's constructions leads to an approach by semidefinite
programming in the context of multiple interactive proof systems with entanglement.
Semidefinite programming allows for an efficient way to approximate Tsirelson's 
bounds for any CHSH-type inequalities numerically. However, it can also be used to prove Tsirelson type bounds
analytically. As an illustration, we first give an alternative proof of Tsirelson's original bound using
semidefinite programming. We then prove a new Tsirelson's bound for the following 
generalized CHSH inequality~\cite{peres:book,braunstein:inequ}. Classically, it can be shown that
\bSpace
|\sum_{i = 1}^n \ev{X_i}{Y_i} + \sum_{i = 1}^{n-1} \ev{X_{i+1}}{Y_i} - \ev{X_1}{Y_n}| \leq 2n - 2.
\eSpace 
Here, we show that for quantum mechanics
\bSpace
|\sum_{i = 1}^n \ev{X_i}{Y_i} + \sum_{i = 1}^{n-1} \ev{X_{i+1}}{Y_i} - \ev{X_1}{Y_n}| 
\leq 2 n\cos\left(\frac{\pi}{2n}\right),
\eSpace
where $\{X_1,\ldots,X_n\}$ and $\{Y_1,\ldots,Y_n\}$ are observables with eigenvalues $\pm 1$
employed by Alice and Bob respectively, corresponding to their $n$ possible measurement settings.
It is well known that this bound can be achieved~\cite{peres:book,braunstein:inequ} for a specific 
set of measurement settings if Alice and Bob share a singlet state. Here, we show that this bound
is indeed optimal for \emph{any} state $\ket{\Psi}$ and choice of measurement settings. This method
generalizes to other CHSH inequalities, for example, the inequality 
considered by Gisin~\cite{gisin:chsh}. As outlined below, Tsirelson's results also imply 
that any bound proved using this method can indeed be achieved using quantum mechanics. 
As Braunstein and Caves~\cite{braunstein:inequ} have shown, it is interesting to consider
inequalities based on many measurement settings, in particular, the chained CHSH inequality above.
The gap between the classical and the quantum bound 
for this inequality is larger than for the original CHSH inequality 
with only two measurement settings. 
They show that even for real experiments that inevitably include noise, this inequality
leads to a stronger violation of local realism, and may thus lead to a better test. 

\section{Preliminaries}

Throughout this \paper, we write $u = (u_1,\ldots,u_n)$ for an $n$-element vector.
$u \cdot v$ denotes the inner product between vectors $u$ and $v$.
Furthermore, $\diag(\lambda)$ denotes the matrix with 
the components of the vector $\lambda$
on its diagonal. 
We write $A = [a_{ij}]$
to indicate that the entry in the $i$-th row and $j$-th column of $A$ is $a_{ij}$.
We also use the shorthand $[n] = \{1,\ldots,n\}$.
$A^{\dagger}$ is the conjugate transpose of matrix $A$. 
A positive semidefinite $n \times n$ 
matrix $A$ is a nonsingular Hermitian matrix such that $x^*Ax \geq 0$ for all $x \in \Complex^n$~\cite{horn&johnson:ma}. 
We use $A \succeq 0$ to indicate that $A$ is positive semidefinite. 
It will be important that a Hermitian matrix $A$ is positive semidefinite if and only if 
all of its eigenvalues are nonnegative~\cite[Theorem 7.2.1]{horn&johnson:ma}.

We will need two ingredients for our proof. 
First, the following result
by Tsirelson~\cite[Theorem 1]{tsirel:original} and~\cite{tsirel:separated,tsirel:hadron} plays an essential role.
\begin{theorem}[Tsirelson]\label{tsirel}
Let $X_1,\ldots,X_n$ and $Y_1,\ldots,Y_n$ be observables with eigenvalues in the interval $[-1;1]$.
Then for any state $\ket{\Psi} \in \mA \otimes \mB$ and for all $s,t \in [n]$
there exist real unit vectors $x_1,\ldots,x_n$,$y_1,\ldots,y_n \in \Real^{2n}$ such that
\bSpace
\bra{\Psi}X_s \otimes Y_t\ket{\Psi} = x_s \cdot y_t.
\eSpace
Conversely, let $x_s,y_t \in \Real^{N}$ be real unit vectors.
Let $\ket{\Psi} \in \mA \otimes \mB$ be any maximally entangled state where 
$\dim(\mA) = \dim(\mB) = 2^{\ceil{N/2}}$. Then for all $s,t \in [n]$
there exist observables $X_s$ on $\mA$ and $Y_t$ on $\mB$ with eigenvalues $\pm 1$
such that
\bSpace
x_s \cdot y_t = \bra{\Psi}X_s \otimes Y_t\ket{\Psi}.
\eSpace
\end{theorem}
In particular, this means that we can rewrite CHSH inequalities in terms of vectors.
The second part of Tsirelson's result implies that any strategy based on 
vectors can indeed be implemented using quantum measurements. 
See~\cite{tsirel:hadron} for a detailed construction.

Secondly, we will make use of semidefinite programming. This is a special 
case of convex optimization. We refer to~\cite{boyd:book} for an in-depth introduction.
The goal of semidefinite programming is to solve he following semidefinite program (SDP)
in terms of the variable $X \in S^n$
\begin{sdp}{maximize}{$\Tr(CX)$}
&$\Tr(A_iX) = b_i, i = 1,\ldots,p$,
and $X \succeq 0$
\end{sdp}
for given matrices $C,A_1,\ldots,A_p \in S^n$ where $S^n$ is the space of symmetric
$n \times n$ matrices. 
$X$ is called \emph{feasible}, if it satisfies all constraints.
An important aspect of 
semidefinite programming is duality.  Intuitively, the idea behind Lagrangian duality is to extend the objective function
(here $\Tr(CX)$) with a weighted sum of the constraints in such a way, that we will be penalized
if the constraints are not fulfilled. The weights then correspond to the dual variables. 
Optimizing over these weights then gives rise to the \emph{dual problem}.
The original problem is called the \emph{primal problem}.
An example of this approach is given in the next section.
Let $d'$ denote the optimal value of the dual problem, and $p'$ the optimal value
of the primal problem from above. Weak duality says that $d' \geq p'$. 
In particular, if we have $d' = p'$ for a feasible dual and primal solution
respectively, we can conclude that both solutions are optimal. 

\section{Tsirelson's bound}\label{tsirelBoundSection}

To illustrate our approach we first give a detailed proof of Tsirelson's bound 
using semidefinite programming. This proof is more complicated than Tsirelson's original 
proof, however, it serves as a good introduction to the following section.
Let $X_1,X_2$ and $Y_1,Y_2$ denote the observables with eigenvalues $\pm 1$
used by Alice and Bob respectively. Our goal is now to show an upper bound for
\bSpace
|\ev{X_1}{Y_1} + \ev{X_1}{Y_2} + \ev{X_2}{Y_1} - \ev{X_2}{Y_2}|.
\eSpace
From Theorem~\ref{tsirel} we know that there exist
real unit vectors $x_s,y_t \in \Real^4$ such that for all $s,t \in \01$
$\ev{X_s}{Y_t} = x_s \cdot y_t$. 
In order to find Tsirelson's bound, we thus want to solve the following
problem: 
maximize $x_1 \cdot y_1 + x_1 \cdot y_2 + x_2 \cdot y_1 - x_2 \cdot y_2$, 
subject to $\norm{x_1} = \norm{x_2} = \norm{y_1} = \norm{y_2} = 1$. Note that we can
drop the absolute value since any set of vectors maximizing the above equation, simultaneously
leads to a set of vectors minimizing it by taking $-y_1,-y_2$ instead.
We will now phrase this as a semidefinite program. Let $G = [g_{ij}]$ be the 
Gram matrix of the vectors $\{x_1,x_2,y_1,y_2\} \subseteq \Real^4$ with respect to the inner product:
$$
G = 
\left(\begin{array}{cccc}
x_1 \cdot x_1 & x_1 \cdot x_2 & x_1 \cdot y_1 & x_1 \cdot y_2 \\
x_2 \cdot x_1 & x_2 \cdot x_2 & x_2 \cdot y_1 & x_2 \cdot y_2\\
y_1 \cdot x_1 & y_1 \cdot x_2 & y_1 \cdot y_2 & y_1 \cdot y_2\\
y_2 \cdot x_1 & y_2 \cdot x_2 & y_2 \cdot y_1 & y_2 \cdot y_2
\end{array}\right).
$$
$G$ can thus be written as $G = B^TB$ where the columns of $B$ are the vectors
$\{x_1,x_2,y_1,y_2\}$. By~\cite[Theorem 7.2.11]{horn&johnson:ma} we can write $G = B^TB$ if
and only if $G$ is positive semidefinite.
We thus impose the constraint that $G \succeq 0$.
To make sure that we obtain unit vectors, we add the constraint that
all diagonal entries of $G$ must be equal to $1$.
Define
$$
W = \left(\begin{array}{cccc} 0 &0& 1& 1\\
0  & 0 & 1 & -1\\
1 & 1 & 0 & 0\\
1 & -1 & 0 & 0
\end{array}\right).
$$
Note that the choice of order of the vectors in $B$ is not unique, however, a different order
only leads to a different $W$ and does not change our argument.
We can now
rephrase our optimization problem as the following SDP:
\begin{sdp}{maximize}{$\frac{1}{2}\Tr(GW)$}
&$G \succeq 0$
and $\forall i, g_{ii} = 1$
\end{sdp}
We can then write for the Lagrangian 
\bSpace
L(G,\lambda) = \frac{1}{2}\Tr(GW) - \Tr(\diag(\lambda)(G - I)),
\eSpace
where $\lambda = (\lambda_1, \lambda_2, \lambda_3,\lambda_4)$.
The dual function is then
\begin{eqnarray*}
g(\lambda) &=& \sup_G \Tr\left(G\left(\frac{1}{2}W - \diag(\lambda)\right)\right) + \Tr(\diag(\lambda))\\
           &=& \left\{\begin{array}{ll}
                       \Tr(\diag(\lambda)) & \mbox{if } \frac{1}{2}W - \diag(\lambda) \preceq 0\\[0.5mm]
                       \infty & \mbox{otherwise}
                      \end{array} \right.
\end{eqnarray*}
We then obtain the following dual formulation of the SDP
\begin{sdp}{minimize}{$\Tr(\diag(\lambda))$}
&$-\frac{1}{2}W + \diag(\lambda) \succeq 0$
\end{sdp}
Let $p'$ and $d'$ denote optimal values for the primal and Lagrange dual 
problem respectively. From weak duality it follows that $d' \geq p'$. 
For our example, it is not difficult to see that this is indeed true. Let $G'$ and $\lambda'$
be optimal solutions of the primal and dual problem, i.e. $p' = \frac{1}{2}\Tr(G'W)$ and 
$d' = \Tr(\diag(\lambda'))$. Recall that all entries on the diagonal of $G'$ are 1
and thus $\Tr(\diag(\lambda')) = \Tr(G'\diag(\lambda'))$.
Then 
$d' - p' = \Tr(\diag(\lambda')) - \frac{1}{2}\Tr(G'W) = \Tr(G'\diag(\lambda') - \frac{1}{2}G'W) = \Tr(G'(\diag(\lambda') - \frac{1}{2}W)) \geq 0$,
where the last inequality follows from the constraints $G' \succeq 0$, $\diag(\lambda') - \frac{1}{2}W \succeq 0$
and~\cite[Example 2.24]{boyd:book}.

In order to prove Tsirelson's bound, we will now exhibit an optimal solution for both the primal
and dual problem and then show that the value of the primal problem equals the value of the dual problem.
The optimal solution is well 
known~\cite{tsirel:original,tsirel:separated,peres:book}.
Alternatively, we could easily guess the optimal solution based on numerical 
optimization by a small program for
Matlab~\footnote{See {http://www.cwi.nl/\~{}wehner/tsirel/} for the Matlab example code.} and the package 
SeDuMi~\cite{sedumi} for semidefinite programming. Consider the following solution for the
primal problem
$$
G' = \left( \begin{array}{cccc}
              1 & 0 & \frac{1}{\sqrt{2}} & \frac{1}{\sqrt{2}}\\
              0 & 1 & \frac{1}{\sqrt{2}} & -\frac{1}{\sqrt{2}}\\
              \frac{1}{\sqrt{2}} & \frac{1}{\sqrt{2}} & 1 & 0\\
              \frac{1}{\sqrt{2}} & - \frac{1}{\sqrt{2}} & 0 & 1\\
            \end{array}
     \right),
$$
which gives rise to the primal value $p' = \frac{1}{2}\Tr(G'W) = 2\sqrt{2}$.
Note that $G' \succeq 0$ since all its eigenvalues are nonnegative~\cite[Theorem 7.2.1]{horn&johnson:ma}
and all its diagonal entries are 1. Thus all constraints are satisfied.
The lower left quadrant of $G'$ is in fact the same as the well known 
correlation matrix for 2 observables~\cite[Equation 3.16]{tsirel:hadron}.
Next, consider the following solution for the dual problem
\bSpace
\lambda' = \frac{1}{\sqrt{2}} \left(1,1,1,1\right).
\eSpace
The dual value is then $d' = \Tr(\diag(\lambda')) = 2\sqrt{2}$.
Because $-W + \diag(\lambda') \succeq 0$, $\lambda'$ satisfies the constraint.
Since $p' = d'$, $G'$ and $\lambda'$ are in fact optimal solutions for
the primal and dual respectively. We can thus conclude that
\bSpace
|\ev{X_1}{Y_1} + \ev{X_1}{Y_2} + \ev{X_2}{Y_1} - \ev{X_2}{Y_2}| \leq 2\sqrt{2},
\eSpace
which is Tsirelson's bound~\cite{tsirel:original}. By Theorem~\ref{tsirel}, this
bound is achievable.

\section{Tsirelson's bounds for more than 2 observables}\label{generalSemiDef}

We now show how to obtain bounds for inequalities based on more than 2 observables
for both Alice and Bob. In particular, we will prove a bound for the chained
CHSH inequality 
for the quantum case.
It is well known~\cite{peres:book} that it is possible to choose observables
$X_1,\ldots,X_n$ and $Y_1,\ldots,Y_n$ such that
\bSpace
|\sum_{i = 1}^n \ev{X_i}{Y_i} + \sum_{i = 1}^{n-1} \ev{X_{i+1}}{Y_i} - \ev{X_1}{Y_n}|
= 2 n \cos\left(\frac{\pi}{2n}\right).
\eSpace
We now show that this is optimal. Our proof is similar to the last section.
However, it is more difficult to show feasibility for
all $n$.
\begin{theorem}
Let $\rho \in \mA \otimes \mB$ be an arbitrary state, where $\mA$ and $\mB$
denote the Hilbert spaces of Alice and Bob. Let $X_1,\ldots,X_n$ and $Y_1,\ldots,Y_n$ 
be observables with eigenvalues $\pm 1$ on $\mA$ and $\mB$ respectively. 
Then 
$$
|\sum_{i = 1}^n \ev{X_i}{Y_i} + \sum_{i = 1}^{n-1} \ev{X_{i+1}}{Y_i} - \ev{X_1}{Y_n}| 
\leq 2 n\cos\left(\frac{\pi}{2n}\right),
$$
\end{theorem}

\begin{proof}
By Theorem~\ref{tsirel}, our goal is to find the maximum value for
$
x_1 \cdot y_1 + x_2 \cdot y_1 + x_2 \cdot y_2 + x_3 \cdot y_2 + \ldots
+ x_n \cdot y_n - x_1 \cdot y_n,
$
for real unit vectors $x_1,\ldots,x_n,y_1,\ldots,y_n \in \Real^{2n}$. 
As above we can drop the absolute value. 
Let $G = [g_{ij}]$ be the Gram matrix of the vectors 
$\{x_1,\ldots,x_n,y_1,\ldots,y_n\} \subseteq \Real^{2n}$.
As before, we can thus write $G = B^TB$, where the 
columns of $B$ are the vectors $\{x_1,\ldots,x_n,y_1,\ldots,y_n\}$, if
and only if $G\succeq 0$.
To ensure we obtain unit vectors, we again demand that 
all diagonal entries of $G$ equal $1$. 
Define $n \times n$ matrix $A$ and $2n \times 2n$ matrix $W$ by

\begin{eqnarray*}
A = \left(\begin{array}{ccccc}
            1 & 1 & 0 & \ldots &  0\\
            0 & 1 & 1 &  &  \vdots \\
            \vdots &  & \ddots & \ddots &  0\\
            0  &  & &               1 & 1\\        
            -1 & 0 & \ldots &  0 & 1\\
          \end{array}
    \right),\mbox{~}
W = \left(\begin{array}{cc} 
            0 & A^{\dagger}\\
            A & 0 
          \end{array}
     \right).
\end{eqnarray*}
We can now phrase our maximization problem as the following SDP:
\begin{sdp}{maximize}{$\frac{1}{2}\Tr(GW)$}
&$G \succeq 0$
and $\forall i, g_{ii} = 1$
\end{sdp}
Analog to the previous section,
the dual SDP is then:
\begin{sdp}{minimize}{$\Tr(\diag(\lambda))$}
&$-\frac{1}{2}W + \diag(\lambda) \succeq 0$
\end{sdp}
Let $p'$ and $d'$ denote optimal values for the primal and dual 
problem respectively. As before, $d' \geq p'$. 

\paragraph{Primal}
We will now show
that the vectors suggested in~\cite{peres:book} are optimal. For $k \in [n]$, choose 
unit vectors $x_k,y_k \in \Real^{2n}$ to be of the form
\begin{eqnarray*}
x_k &=& (\cos(\phi_k), \sin(\phi_k),0,\ldots,0),\\
y_k &=& (\cos(\psi_k), \sin(\psi_k),0,\ldots,0),
\end{eqnarray*}
where $\phi_k = \frac{\pi}{2n}(2k - 2)$ and $\psi_k = \frac{\pi}{2n}(2k - 1)$.
The angle between $x_k$ and $y_k$ is given by $\psi_k - \phi_k = \frac{\pi}{2n}$
and thus $x_k \cdot y_k = \cos\left(\frac{\pi}{2n}\right)$.
The angle between $x_{k+1}$ and $y_k$ is $\phi_{k+1} - \psi_k = \frac{\pi}{2n}$
and thus $x_{k+1} \cdot y_k = \cos\left(\frac{\pi}{2n}\right)$. 
Finally, the angle between
$-x_1$ and
$y_n$
is $\pi - \psi_n = \frac{\pi}{2n}$ and so $-x_1 \cdot y_n = \cos\left(\frac{\pi}{2n}\right)$.
The value of our primal problem is thus given by
\bSpace
p' = \sum_{k=1}^n x_k \cdot y_k + \sum_{k=1}^{n-1} x_{k+1} \cdot y_k - x_1 \cdot y_{n}
= 2n \cos\left(\frac{\pi}{2n}\right).
\eSpace 
Let $G'$ be the Gram matrix constructed from all vectors $x_k,y_k$ as described earlier.
Note that our constraints are satisfied: $\forall i: g_{ii} = 1$ and $G' \succeq 0$, because
$G'$ is symmetric and of the form $G' = B^TB$.

\paragraph{Dual}
Now consider the $n$-dimensional vector
\bSpace
\lambda' = \cos\left(\frac{\pi}{2n}\right) \left(1,\ldots,1\right).
\eSpace
In order to show that this is a feasible solution to the dual problem, we have to prove
that $-\frac{1}{2}W + \diag(\lambda') \succeq 0$ and thus the
constraint is satisfied. 
To this end, we first show that
\begin{claim}\label{aClaim}
The eigenvalues of $A$ are given by
$\gamma_s = 1 + e^{i\pi(2s+1)/n}$ with $s = 0,\ldots,n-1$.
\end{claim}
\begin{proof}
Note that if the lower left corner of $A$ were $1$, $A$ would
be a circulant matrix~\cite{circulant}, i.e.~each row of $A$ 
is constructed by taking the previous row and shifting it one place to the right.
We can use ideas from
circulant matrices to guess eigenvalues $\gamma_s$ with eigenvectors
\bSpace
u_s = (\rho_s^{n-1},\rho_s^{n-2},\rho_s^{n-3},\ldots,\rho_s^1,\rho_s^0),
\eSpace 
where $\rho_s = e^{-i\pi(2s+1)/n}$ and $s = 0,\ldots,n-1$. 
By definition, $u = (u_1,u_2,\ldots,u_n)$ is an eigenvector of $A$ with 
eigenvalue $\gamma$ if and only if $Au = \gamma u$. Here, $Au = \gamma u$
if and only if 
\begin{eqnarray*}
(i)&& 
\forall j \in \{1,\ldots,n-1\}:
u_j + u_{j+1} = \gamma u_j,\\
(ii)&& -u_1 + u_n = \gamma u_n.
\end{eqnarray*}
Since for any $j \in \{1,\ldots,n-1\}$
\begin{eqnarray*}
u_j + u_{j+1} 
&=&
\rho_s^{n-j} + \rho_s^{n-j-1} = 
\\
&=&
e^{-i(n-j)\pi(2s+1)/n}(1+e^{i\pi(2s+1)/n}) = \\
&=&
\rho_s^{n-j} \gamma_s = \gamma_s u_j,
\end{eqnarray*}
(i) is satisfied.
Furthermore 
(ii) is satisfied, since
\begin{eqnarray*}
-u_1 + u_n &=& - \rho_s^{n-1} + \rho_s^0 = \\
&=& 
- e^{-i \pi(2s+1)} e^{i\pi(2s+1)/n} + 1 = \\
&=&
1 + e^{i\pi(2s+1)/n} = \\
&=&
\gamma_s \rho_s^0 = \gamma_s u_n.
\end{eqnarray*}
\end{proof}

\begin{claim}\label{biggestVal}
The largest eigenvalue of $W$ is given by $\gamma = 2\cos\left(\frac{\pi}{2n}\right)$.
\end{claim}
\begin{proof}
By~\cite[Theorem 7.3.7]{horn&johnson:ma}, the eigenvalues of $W$ are given
by the singular values of $A$ and their negatives. It follows from
Claim~\ref{aClaim} 
that the singular values of $A$ are
\bSpace
\sigma_s = \sqrt{\gamma_s \gamma_s^*} = 2 + 2\cos\left(\frac{\pi(2s+1)}{n}\right).
\eSpace 
Considering the shape of the cosine function, it is easy to see that the largest 
singular value of $A$ is given by $2 + 2\cos(\pi/n)
= 4 \cos^2(\pi/(2n))$,
the largest eigenvalue of $W$ is 
$\sqrt{2 + 2\cos(\pi/n)} =  2 \cos(\pi/(2n))$.
\end{proof}

Since $-\frac{1}{2}W$ and $\diag(\lambda')$ are both Hermitian, Weyl's 
theorem~\cite[Theorem 4.3.1]{horn&johnson:ma} implies that 
\begin{eqnarray*}
&&\gamma_{min}\left(-\frac{1}{2}W + \diag(\lambda')\right)\\
&\geq&
\gamma_{min}\left(-\frac{1}{2}W\right) + \gamma_{min}\left(\diag(\lambda')\right),
\end{eqnarray*}
where $\gamma_{min}(M)$ is the smallest eigenvalue of a matrix $M$. It then follows 
from the fact that $\diag(\lambda')$ is diagonal and 
Claim~\ref{biggestVal} 
that
\begin{eqnarray*}
&&\gamma_{min}\left(-\frac{1}{2}W + \diag(\lambda')\right)\\
 &\geq& 
-\frac{1}{2}\left(2\cos\left(\frac{\pi}{2n}\right)\right) + \cos\left(\frac{\pi}{2n}\right)
= 
0.
\end{eqnarray*}
Thus $-\frac{1}{2}W + \diag(\lambda') \succeq 0$
and $\lambda'$ is a feasible solution to the dual problem. The value of the
dual problem is then
\bSpace
d' = \Tr(\diag(\lambda')) = 2 n \cos\left(\frac{\pi}{2n}\right).
\eSpace 
Because $p' = d'$, $G'$ and $\lambda'$ are optimal solutions for
the primal and dual respectively,
which completes our proof.
\end{proof}

Note that for the primal problem we are effectively dealing with $2$-dimensional 
vectors, $x_k,y_k$. It therefore follows from
Tsirelson's construction~\cite{tsirel:hadron} that given an EPR pair we can find observables
such that the bound is tight. In fact, these vectors just determine the measurement directions
as given in~\cite{peres:book}.

\section{Discussion}
Our approach can be generalized to other CHSH-type inequalities.
For another inequality, we merely use a different matrix $A$ in $W$. For example, for Gisin's CHSH 
inequality~\cite{gisin:chsh}, $A$ is the matrix with 1's in the upper left half and on the diagonal, and
-1's in the lower right part. Otherwise our approach stays exactly the same, and thus we do not consider 
this case here. Numerical results provided by our Matlab example code suggest that Gisin's observables
are optimal.
Given the framework of semidefinite
programming, the only difficulty in proving bounds for other inequalities is to determine the eigenvalues
of the corresponding $A$, a simple matrix.
Since our approach is based on Tsirelson's vector construction, it is limited to CHSH-type inequalities
for two parties.

\section{Acknowledgments}

Many thanks to Boris Tsirelson for sending me a copy of~\cite{tsirel:original}
and~\cite{tsirel:separated}, to Oded Regev for pointing me to~\cite{boyd:book}
and for introducing me to the concept of approximation algorithms 
such as~\cite{goemans:maxcut}, to Serge Massar for pointers, to Ronald de Wolf for the pointer
to~\cite{circulant} and proofreading, and to Manuel Ballester for alerting me to correct the journal reference of~\cite{braunstein:inequ}. Supported by EU project
RESQ IST-2001-37559 and NWO vici project 2004-2009.


\begin{thebibliography}{10}

\bibitem{bell:epr}
J.~S. Bell.
\newblock On the {E}instein-{P}odolsky-{R}osen paradox.
\newblock {\em Physics}, 1:195--200, 1965.

\bibitem{boyd:book}
S.~Boyd and L.~Vandenberghe.
\newblock {\em Convex Optimization}.
\newblock Cambridge University Press, 2004.

\bibitem{braunstein:inequ}
S.L. Braunstein and C.M. Caves.
\newblock Wringing out better {B}ell inequalities.
\newblock {\em Annals of Physics}, 202:22--56, 1990.

\bibitem{massar:tsirel}
H.~Buhrman and S.~Massar.
\newblock Causality and {C}irel'son bounds.
\newblock quant-ph/0409066, 2004.

\bibitem{chsh:nonlocal}
J.~Clauser, M.~Horne, A.~Shimony, and R.~Holt.
\newblock Proposed experiment to test local hidden-variable theories.
\newblock {\em Physical Review Letters}, 23:880--884, 1969.

\bibitem{cleve:nonlocal}
R.~Cleve, P.~H{\o}yer, B.~Toner, and J.~Watrous.
\newblock Consequences and limits of nonlocal strategies.
\newblock In {\em Proceedings of 19th IEEE Conference on Computational
  Complexity}, pages 236--249, 2004.
\newblock quant-ph/0404076.

\bibitem{filipp:bell}
S.~Filipp and K.~Svozil.
\newblock Tracing the bounds on {B}ell-type inequalities.
\newblock In {\em Proceedings of Foundations of Probability and Physics-3},
  pages 87--94, 2004.

\bibitem{gisin:chsh}
N.~Gisin.
\newblock {B}ell inequality for arbitrary many settings of the analyzers.
\newblock {\em Physics Letters A}, 260:1--3, 1999.

\bibitem{goemans:maxcut}
M.X. Goemans and D.P. Williamson.
\newblock Improved approximation algorithms for maximum cut and satisfiability
  problems using semidefinite programming.
\newblock {\em J. Assoc. Comput. Mach.}, 42:1115--1145, 1995.

\bibitem{horn&johnson:ma}
R.~A. Horn and C.~R. Johnson.
\newblock {\em Matrix Analysis}.
\newblock Cambridge University Press, 1985.

\bibitem{peres:book}
A.~Peres.
\newblock {\em Quantum Theory: Concepts and Methods}.
\newblock Kluwer Academic Publishers, 1993.

\bibitem{peres:allBell}
A.~Peres.
\newblock All the {B}ell inequalities.
\newblock {\em Foundations of Physics}, 29:589--614, 1999.

\bibitem{circulant}
R.M.Gray.
\newblock {\em Toeplitz and Circulant Matrices: A review}.
\newblock 1971.

\bibitem{sedumi}
J.~Sturm and AdvOL.
\newblock {S}e{D}u{M}i.
\newblock http://sedumi.mcmaster.ca/.

\bibitem{tsirel:separated}
B.~Tsirelson.
\newblock Quantum analogues of {B}ell inequalities: The case of two spatially
  separated domains.
\newblock {\em Journal of Soviet Mathematics}, 36:557--570, 1987.

\bibitem{tsirel:hadron}
B.~Tsirelson.
\newblock Some results and problems on quantum {B}ell-type inequalities.
\newblock {\em Hadronic Journal Supplement}, 8(4):329--345, 1993.

\bibitem{tsirel:original}
B.~Cirel'son (Tsirelson).
\newblock Quantum generalizations of {B}ell's inequality.
\newblock {\em Letters in Mathematical Physics}, 4:93--100, 1980.

\end{thebibliography}
\end{document}